\documentclass[journal]{IEEEtran}

\usepackage{graphicx}  

\usepackage{amsmath}   

\usepackage{epsfig}
\newtheorem{theorem}{\indent Theorem}

\newtheorem{definition}[theorem]{\indent Definition}

\begin{document}
%
\title{Quantum Reinforcement Learning}
%
%
\author{Daoyi~Dong,
        Chunlin~Chen,
        Hanxiong Li,
        Tzyh-Jong~Tarn
\thanks{This work was supported in part by the National Natural Science
Foundation of China under Grant No. 60703083, the K. C. Wong
Education Foundation (Hong Kong), the China Postdoctoral Science
Foundation (20060400515) and by a grant from RGC of Hong Kong
(CityU:116406).}
\thanks{D. Dong is with the Key Laboratory of Systems and
Control, Institute of Systems Science, AMSS, Chinese Academy of
Sciences, Beijing 100190, China (email: dydong@amss.ac.cn).}
\thanks{C. Chen is with the Department of Control and System Engineering, Nanjing
University, Nanjing 210093, China (email: clchen@nju.edu.cn).}
\thanks{H. Li is with the Central South University, Changsha 410083, China
and the Department of Manufacturing Engineering and Engineering
Management, City University of Hong Kong, Hong Kong.}
\thanks{T. J. Tarn is with the Department of Electrical and
Systems Engineering, Washington University in St. Louis, St.
Louis, MO 63130 USA.}}
\maketitle

\begin{abstract}
The key approaches for machine learning, especially learning in
unknown probabilistic environments are new representations and
computation mechanisms. In this paper, a novel quantum
reinforcement learning (QRL) method is proposed by combining
quantum theory and reinforcement learning (RL). Inspired by the
state superposition principle and quantum parallelism, a framework
of value updating algorithm is introduced. The state (action) in
traditional RL is identified as the eigen state (eigen action) in
QRL. The state (action) set can be represented with a quantum
superposition state and the eigen state (eigen action) can be
obtained by randomly observing the simulated quantum state
according to the collapse postulate of quantum measurement. The
probability of the eigen action is determined by the probability
amplitude, which is parallelly updated according to rewards. Some
related characteristics of QRL such as convergence, optimality and
balancing between exploration and exploitation are also analyzed,
which shows that this approach makes a good tradeoff between
exploration and exploitation using the probability amplitude and
can speed up learning through the quantum parallelism. To evaluate
the performance and practicability of QRL, several simulated
experiments are given and the results demonstrate the
effectiveness and superiority of QRL algorithm for some complex
problems. The present work is also an effective exploration on the
application of quantum computation to artificial intelligence.
\end{abstract}


%
\IEEEpeerreviewmaketitle

\section{Introduction}
\PARstart{L}{earning} methods are generally classified into
supervised, unsupervised and \emph{reinforcement learning} (RL).
Supervised learning requires explicit feedback provided by
input-output pairs and gives a map from inputs to outputs.
Unsupervised learning only processes on the input data. In
contrast, RL uses a scalar value named \emph{reward} to evaluate
the input-output pairs and learns a mapping from \emph{states} to
\emph{actions} by interaction with the environment through
trial-and-error. Since 1980s, RL has become an important approach
to machine learning \cite{Sutton and Barto 1998}-\cite{Kaya and
Alhajj 2005}, and is widely used in artificial intelligence,
especially in robotics \cite{Beom and Cho 1995}-\cite{Kondo and
Ito 2004}, \cite{Tzafestas and Rigatos 2002}, due to its good
performance of on-line adaptation and powerful learning ability to
complex nonlinear systems. However there are still some difficult
problems in practical applications. One problem is the exploration
strategy, which contributes a lot to better balancing of
\emph{exploration} (trying previously unexplored strategies to
find better policy) and \emph{exploitation} (taking the most
advantage of the experienced knowledge). The other is its slow
learning speed, especially for the complex problems sometimes
known as ``the curse of dimensionality" when the state-action
space becomes huge and the number of parameters to be learned
grows exponentially with its dimension.

To combat those problems, many methods have been proposed in
recent years. Temporal abstraction and decomposition methods have
been explored to solve such problems as RL and dynamic programming
(DP) to speed up learning \cite{Wiering and Schmidhuber
1997}-\cite{Theocharous 2002}. Different kinds of learning
paradigms are combined to optimize RL. For example, Smith
\cite{Smith 2002} presented a new model for representation and
generalization in model-less RL based on the self-organizing map
(SOM) and standard Q-learning. The adaptation of Watkins'
Q-learning with fuzzy inference systems for problems with large
state-action spaces or with continuous state spaces is also
proposed \cite{Vengerov et al 2005}, \cite{Glorennec and Jouffe
1997}, \cite{Tzafestas and Rigatos 2002}, \cite{Er and Deng 2004}.
Many specific improvements are also implemented to modify related
RL methods in practice \cite{Beom and Cho 1995}, \cite{Smart and
Kaelbling 2002}, \cite{Kondo and Ito 2004}, \cite{Chen et al
2008}, \cite{Whiteson and Stone 2006}, \cite{Kaya and Alhajj
2005}. In spite of all these attempts more work is needed to
achieve satisfactory successes and new ideas are necessary to
explore more effective representation methods and learning
mechanisms. In this paper, we explore to overcome some
difficulties in RL using quantum theory and propose a novel
quantum reinforcement learning method.

Quantum information processing is a rapidly developing field. Some
results have shown that quantum computation can more efficiently
solve some difficult problems than the classical counterpart. Two
important quantum algorithms, the Shor algorithm \cite{Shor 1994},
\cite{Ekert and Jozsa 1996} and the Grover algorithm \cite{Grover
1996}, \cite{Grover 1997} have been proposed in 1994 and 1996,
respectively. The Shor algorithm can give an exponential speedup
for factoring large integers into prime numbers and it has been
realized \cite{Vandersypen et al 2001} for the factorization of
integer 15 using nuclear magnetic resonance (NMR). The Grover
algorithm can achieve a square speedup over classical algorithms
in unsorted database searching and its experimental
implementations have also been demonstrated using NMR \cite{Chuang
et al 1998}-\cite{Jones et al 1998} and quantum optics \cite{Kwiat
et al 2000}, \cite{Scully and Zubairy 2001} for a system with four
states. Some methods have also been explored to connect quantum
computation and machine learning. For example, the quantum
computing version of artificial neural network has been studied
from the pure theory to the simple simulated and experimental
implementation \cite{Ventura and Martinez 2000}-\cite{Behrman et
al 2000}. Rigatos and Tzafestas \cite{Rigatos and Tzafestas 2002}
used quantum computation for the parallelization of a fuzzy logic
control algorithm to speed up the fuzzy inference. Quantum or
quantum-inspired evolutionary algorithms have been proposed to
improve the existing evolutionary algorithms \cite{Sahin et al
2005}. Hogg and Portnov \cite{Hogg and Portnov 2000} presented a
quantum algorithm for combinatorial optimization of
overconstrained satisfiability (SAT) and asymmetric travelling
salesman (ATSP). Recently the quantum search technique has been
used to dynamic programming \cite{Naguleswaran and White 2005}.
Taking advantage of quantum computation, some novel algorithms
inspired by quantum characteristics will not only improve the
performance of existing algorithms on traditional computers, but
also promote the development of related research areas such as
quantum computers and machine learning. Considering the essence of
computation and algorithms, Dong and his co-workers \cite{Dong et
al 2005} have presented the concept of \emph{quantum reinforcement
learning} (QRL) inspired by the state superposition principle and
quantum parallelism. Following this concept, we in this paper give
a formal quantum reinforcement learning algorithm framework and
specifically demonstrate the advantages of QRL for speeding up
learning and obtaining a good tradeoff between exploration and
exploitation of RL through simulated experiments and some related
discussions.

This paper is organized as follows. Section II contains the
prerequisite and problem description of standard reinforcement
learning, quantum computation and related quantum gates. In
Section III, quantum reinforcement learning is introduced
systematically, where the state (action) space is represented with
the quantum state, the exploration strategy based on the collapse
postulate is achieved and a novel QRL algorithm is proposed
specifically. Section IV analyzes related characteristics of QRL
such as the convergence, optimality and the balancing between
exploration and exploitation. Section V describes the simulated
experiments and the results demonstrate the effectiveness and
superiority of QRL algorithm. In Section VI, we briefly discuss
some related problems of QRL for future work. Concluding remarks
are given in section VII.

\section{Prerequisite and problem description}
In this section we first briefly review the standard reinforcement
learning algorithms and then introduce the background of quantum
computation and some related quantum gates.

\subsection{Reinforcement learning (RL)}
Standard framework of reinforcement learning is based on
discrete-time, finite-state Markov decision processes (MDPs)
\cite{Sutton and Barto 1998}.

\begin{definition}[MDP]
A Markov decision process (MDP) is composed of the following five
factors:
$\{S,A_{(i)},p_{ij}(a),r_{(i,a)},V,i,j\in{S},a\in{A_{(i)}}\}$,
where: $S$ is the state space; $A_{(i)}$ is the action space for
state $i$; $p_{ij}(a)$ is the probability for state transition;
$r$ is a reward function, $r:\Gamma\to(-\infty,+\infty)$, where
$\Gamma=\{(i,a)|i \in S, a \in A_{(i)}\}$; $V$ is a criterion
function or objective function.
\end{definition}

According to the definition of MDP, we know that the MDP history
is composed of successive states and decisions:
$h_n=(s_0,a_0,s_1,a_1,\dots,s_{n-1},a_{n-1},s_n)$. The policy
$\pi$ is a sequence: $\pi=(\pi_0,\pi_1,\dots)$, when the history
at $n$ is $h_n$, the strategy is adopted to make a decision
according to the probability distribution $\pi_n(\bullet|h_n)$ on
$A_{(s_n)}$.

RL algorithms assume that the state $S$ and action $A_{(s_n)}$ can
be divided into discrete values. At a certain step $t$, the
\emph{agent} observes the state of the environment (inside and
outside of the agent) $s_t$, and then choose an action $a_t$.
After executing the action, the agent receives a reward $r_{t+1}$,
which reflects how good that action is (in a short-term sense).
The state of the environment will change to next state $s_{t+1}$
under the action $a_t$. The agent will choose the next action
$a_{t+1}$ according to related knowledge.

The goal of reinforcement learning is to learn a mapping from
states to actions, that is to say, the agent is to learn a policy
$\pi:S\times \cup_{i\in S}A_{(i)}\to [0,1]$, so that the expected
sum of discounted reward of each state will be maximized:
\begin{eqnarray}
V_{(s)}^{\pi}=E\{r_{(t+1)}+\gamma r_{(t+2)}+\dots|s_t=s,\pi\} \nonumber\\
=E[r_{(t+1)}+\gamma V_{s_{(t+1)}}^{\pi}|s_t=s,\pi]\ \ \ \ \ \ \ \
\nonumber\\
=\sum_{a\in A_s}\pi(s,a)[r_{s}^{a}+\gamma
\sum_{s'}p_{ss'}^{a}V_{(s')}^{\pi}] \ \ \ \ \
\end{eqnarray}
where $\gamma \in [0,1]$ is a discount factor, $\pi(s,a)$ is the
probability of selecting action $a$ according to state $s$ under
policy $\pi$, $p_{ss'}^{a}=Pr\{s_{t+1}=s'|s_t=s,a_t=a\}$ is the
probability for state transition and
$r_{s}^{a}=E\{r_{t+1}|s_t=s,a_t=a\}$ is the expected one-step
reward. $V_{(s)}$ (or $V(s)$) is also called the value function of
state $s$ and the temporal difference (TD) one-step updating rule
of $V(s)$ may be described as
\begin{equation}
V(s) \leftarrow V(s) + \alpha (r+ \gamma V(s') - V(s))
\end{equation}
where $\alpha  \in (0, 1)$ is the learning rate. We have the
optimal state-value function
\begin{equation}\label{Bellman}
V_{(s)}^{*}=\max_{a\in A_s}[r_{s}^{a}+\gamma
\sum_{s'}p_{ss'}^{a}V_{(s')}^{*}]
\end{equation}
\begin{equation}
\pi^{*}=\arg \max_{\pi}V_{(s)}^{\pi},  \forall s\in S
\end{equation}
In dynamic programming, (\ref{Bellman}) is also called the Bellman
equation of $V^{*}$.

As for state-action pairs, there are similar value functions and
Bellman equations, and $Q_{(s,a)}^{\pi}$ stands for the value of
taking the action $a$ in the state $s$ under the policy $\pi$:
\begin{eqnarray}
Q_{(s,a)}^{\pi}=E\{r_{(t+1)}+\gamma r_{(t+2)}+\dots|s_t=s,a_t=a,\pi\} \nonumber\\
=r_{s}^{a}+\gamma\sum_{s'}p_{ss'}^{a}V_{(s')}^{\pi}\ \ \ \ \ \ \ \
\ \ \ \ \ \ \ \ \ \ \ \ \ \ \ \ \ \ \ \ \
\nonumber\\
=r_{s}^{a}+\gamma\sum_{s'}p_{ss'}^{a}\sum_{a'}\pi(s',a')Q_{(s',a')}^{\pi}
\ \ \ \ \ \ \ \ \ \ \
\end{eqnarray}
\begin{equation}
Q_{(s,a)}^{*}=\max_{\pi}Q_{(s,a)}=r_{s}^{a}+\gamma\sum_{s'}p_{ss'}^{a}\max_{a'}Q_{(s',a')}^{\pi}
\end{equation}

Let $\alpha$ be the learning rate, and the one-step updating rule
of Q-learning (a widely used RL algorithm) \cite{Watkins and Dayan
1992} is:
\begin{equation}
Q(s_t,a_t)\leftarrow(1-\alpha)Q(s_t,a_t)+\alpha(r_{t+1}+\gamma
\max_{a'}Q(s_{t+1},a'))
\end{equation}
There are many effective standard RL algorithms like Q-learning,
for example TD($\lambda$), Sarsa, etc. For more details see
\cite{Sutton and Barto 1998}.

\subsection{State superposition and quantum parallelism}
Analogous to classical bits, the fundamental concept in quantum
computation is the \emph{quantum bit} (\emph{qubit}). The two
basic states for a qubit are denoted as $|0\rangle$ and
$|1\rangle$, which correspond to the states 0 and 1 for a
classical bit. However, besides $|0\rangle$ or $|1\rangle$, a
qubit can also lie in the superposition state of $|0\rangle$ and
$|1\rangle$. In other words, a qubit $|\psi\rangle$ can generally
be expressed as a linear combination of $|0\rangle$ and
$|1\rangle$
\begin{equation}\label{qubit}
|\psi\rangle=\alpha|0\rangle+\beta|1\rangle
\end{equation}
where $\alpha$ and $\beta$ are complex coefficients. This special
quantum phenomenon is called \emph{state superposition principle},
which is an important difference between classical computation and
quantum computation \cite{Preskill 1998}.

The physical carrier of a qubit is any two-state quantum system
such as a two-level atom, spin-$\frac{1}{2}$ particle and
polarized photon. For a physical qubit, when we select a set of
bases $|0\rangle$ or $|1\rangle$, we indicate that an observable
$\hat{O}$ of the qubit system has been chosen and the bases
correspond to the two eigenvectors of $\hat{O}$. For convenience,
the measurement process on the observable $\hat{O}$ of a quantum
system in corresponding state $|\psi\rangle$ is directly called a
measurement of quantum state $|\psi\rangle$ in this paper. When we
measure a qubit in superposition state $|\psi\rangle$, the qubit
system would \emph{collapse} into one of its basic states
$|0\rangle$ or $|1\rangle$. However, we cannot determine in
advance whether it will collapse to state $|0\rangle$ or
$|1\rangle$. We only know that we get this qubit in state
$|0\rangle$ with probability $|\alpha|^2$, or in state $|1\rangle$
with probability $|\beta|^2$. Hence $\alpha$ and $\beta$ are
generally called \emph{probability amplitudes}. The magnitude and
argument of probability amplitude represent \emph{amplitude} and
\emph{phase}, respectively. Since the sum of probabilities must be
equal to 1, $\alpha$ and $\beta$ should satisfy
$|\alpha|^2+|\beta|^2=1$.

According to quantum computation theory, a fundamental operation
in the quantum computing process is a unitary transformation $U$
on the qubits. If one applies a transformation $U$ to a
superposition state, the transformation will act on all basis
vectors of this state and the output will be a new superposition
state obtained by superposing the results of all basis vectors. It
seems that the transformation $U$ can simultaneously evaluate the
different values of a function $f(x)$ for a certain input $x$ and
it is called \emph{quantum parallelism}. The quantum parallelism
is one of the most important factors to acquire the powerful
ability of quantum algorithm. However, note that this parallelism
is not immediately useful \cite{Nielsen and Chuang 2000} since the
direct measurement on the output generally gives only $f(x)$ for
one value of $x$. Suppose the input qubit $|z\rangle$ lies in the
superposition state:
\begin{equation}
|z\rangle=\alpha|0\rangle+\beta|1\rangle
\end{equation}
The transformation $U_z$ which describes computing process may be
defined as follows:
\begin{equation}
U_z: |z,0\rangle\rightarrow |z,f(z)\rangle
\end{equation}
where $|z,0\rangle$ represents the joint input state with the
first qubit in $|z\rangle$ and the second qubit in $|0\rangle$,
and $|z,f(z)\rangle$ is the joint output state with the first
qubit in $|z\rangle$ and the second qubit in $|f(z)\rangle$.
According to equations (9) and (10), we can easily obtain
\cite{Nielsen and Chuang 2000}:
\begin{equation}
U_{z}|z,0\rangle=\alpha|0,f(0)\rangle+\beta|1,f(1)\rangle
\end{equation}
The result contains information about both $f(0)$ and $f(1)$, and
we seem to evaluate $f(z)$ for two values of $z$ simultaneously.

The above process corresponds to a ``\emph{quantum black box}" (or
oracle). By feeding quantum superposition states to a quantum
black box, we can learn what is inside with an exponential
speedup, compared to how long it would take if we were only
allowed classical inputs \cite{Preskill 1998}.

Now consider an $n$-qubit system, which can be represented with
tensor product of $n$ qubits:
\begin{equation}
|\phi\rangle=|\psi_1\rangle\otimes|\psi_2\rangle\otimes\dots|\psi_n\rangle=\sum_{x=00\dots0}^{11\dots1}C_x|x\rangle
\end{equation}
where `$\otimes$' means tensor product,
$\sum_{x=00\dots0}^{11\dots1}|C_x|^2=1$, $C_x$ is complex
coefficient and $|C_x|^2$ represents occurrence probability of
$|x\rangle$ when state $|\phi\rangle$ is measured. $x$ can take on
$2^n$ values, so the superposition state can be looked upon as the
superposition of all integers from $0$ to $2^n-1$. Since $U$ is a
unitary transformation, computing function $f(x)$ can result
\cite{Preskill 1998}:
\begin{equation}
U\sum_{x=00\dots0}^{11\dots1}C_x|x,0\rangle=\sum_{x=00\dots0}^{11\dots1}C_xU|x,0\rangle
=\sum_{x=00\dots0}^{11\dots1}C_x|x,f(x)\rangle
\end{equation}

Based on the above analysis, it is easy to find that an $n$-qubit
system can simultaneously process $2^n$ states although only one
of the $2^n$ states is accessible through a direct measurement and
the ability is required to extract information about more than one
value of $f(x)$ from the output superposition state \cite{Nielsen
and Chuang 2000}. This is different from classical parallel
computation, where multiple circuits built to compute are executed
simultaneously, since quantum parallelism doesn't necessarily make
a tradeoff between computation time and needed physical space. In
fact, quantum parallelism employs a single circuit to
simultaneously evaluate the function for multiple values by
exploiting the quantum state superposition principle and provides
an exponential-scale computation space in the $n$-qubit linear
physical space. Therefore quantum computation can effectively
increase the computing speed of some important classical
functions. So it is possible to obtain significant result through
fusing the quantum computation into the reinforcement learning
theory.

\subsection{Quantum Gates}
In the classical computation, the logic operators that can
complete some specific tasks are called \emph{logic gates}, such
as \emph{NOT} gate, \emph{AND} gate, \emph{XOR} gate, and so on.
Analogously, quantum computing tasks can be completed through
\emph{quantum gates}. Nowadays some simple quantum gates such as
quantum NOT gate and quantum \emph{CNOT} gate have been built in
quantum computation. Here we only introduce two important quantum
gates, \emph{Hadamard gate} and \emph{phase gate}, which are
closely related to accomplish some quantum logic operations for
the present quantum reinforcement learning. The detailed
discussion about quantum gates can be found in the Ref.
\cite{Nielsen and Chuang 2000}.

Hadamard gate (or Hadamard transform) is one of the most useful
quantum gates and can be represented as \cite{Nielsen and Chuang
2000}:
\begin{equation}
H\equiv\frac{1}{\sqrt{2}} \left(%
\begin{array}{cc}
  1 & 1 \\
  1 & -1 \\
\end{array}%
\right)
\end{equation}
Through Hadamard gate, a qubit in the state $|0\rangle$ is
transformed into an equally weighted superposition state of
$|0\rangle$ and $|1\rangle$, i.e.
\begin{equation}
H|0\rangle \equiv\frac{1}{\sqrt{2}}\left(%
\begin{array}{cc}
  1 & 1 \\
  1 & -1 \\
\end{array}%
\right)
\left(%
\begin{array}{c}
  1 \\
  0 \\
\end{array}%
\right)=\frac{1}{\sqrt{2}}|0\rangle  + \frac{1}{\sqrt
{2}}|1\rangle
\end{equation}
Similarly, a qubit in the state $|1\rangle$ is transformed into
the superposition state
$\frac{1}{\sqrt{2}}|0\rangle-\frac{1}{\sqrt{2}}|1\rangle$, i.e.
the magnitude of the amplitude in each state is
$\frac{1}{\sqrt{2}}$, but the phase of the amplitude in the state
$|1\rangle$ is inverted. In classical probabilistic algorithms,
the phase has no analog since the amplitudes are in general
complex numbers in quantum mechanics.

The other related quantum gate is the phase gate (conditional
phase shift operation) which is an important element to carry out
the Grover iteration for reinforcing ``good" decision. According
to quantum information theory, this transformation may be
efficiently implemented on a quantum computer. For example, the
transformation describing this for a two-state system is of the
form:
\begin{equation}
U_{phase}=\left(%
\begin{array}{cc}
  1 & 0 \\
  0 & e^{i\varphi} \\
\end{array}%
\right)
\end{equation}
where $i=\sqrt{-1}$ and $\varphi$ is arbitrary real number
\cite{Grover 1997}.

\section{Quantum reinforcement learning (QRL)}
Just like the traditional reinforcement learning, a quantum
reinforcement learning system can also be identified for three
main subelements: a policy, a reward function and a model of the
environment (maybe not explicit). But quantum reinforcement
learning algorithms are remarkably different from all those
traditional RL algorithms in the following intrinsic aspects:
representation, policy, parallelism and updating operation.

\subsection{Representation}
As we represent a QRL system with quantum concepts, similarly, we
have the following definitions and propositions for quantum
reinforcement learning.

\begin{definition}[Eigen states (or eigen actions)]
Select an observable of a quantum system and its eigenvectors form
a set of complete orthonormal bases in a Hilbert space. The states
$s$ (or actions $a$) in Definition 1 are denoted as the
corresponding orthogonal bases and are called the eigen states or
eigen actions in QRL.
\end{definition}

\newtheorem{remark}{Remark}
\begin{remark}
In the remainder of this paper, we indicate that an observable has
been chosen but we do not present the observable specifically when
mentioning a set of orthogonal bases. From Definition 2, we can
get the set of eigen states: $S$, and that of eigen actions for
state $i$: $A_{(i)}$. The eigen state (eigen action) in QRL
corresponds to the state (action) in traditional RL. According to
quantum mechanics, the quantum state for a general closed quantum
system can be represented with a unit vector $|\psi\rangle$ (Dirac
representation) in a Hilbert space. The inner product of
$|\psi_1\rangle$ and $|\psi_2\rangle$ can be written into
$\langle\psi_1|\psi_2\rangle$ and the normalization condition for
$|\psi\rangle$ is $\langle\psi|\psi\rangle=1$. As the simplest
quantum mechanical system, the state of the qubit can be described
as (\ref{qubit}) and its normalization condition is equivalent to
$|\alpha|^2+|\beta|^2=1$.
\end{remark}

\begin{remark}
According to the superposition principle in quantum computation,
since a quantum reinforcement learning system can lie in some
orthogonal quantum states, which correspond to the eigen states
(eigen actions), it can also lie in an arbitrary superposition
state. That is to say, a QRL system which can take on the states
(or actions) {$|\psi_n\rangle$} is also able to occupy their
linear superposition state (or action)
\begin{equation}
|\psi\rangle=\sum_{n}\beta_n|\psi_{n}\rangle
\end{equation}
It is worth noting that this is only a representation method and
our goal is to take advantage of the quantum characteristics in
the learning process. In fact, the state (action) in QRL is not a
practical state (action) and it is only an artificial state
(action) for computing convenience with quantum systems. The
practical state (action) is the eigen state (eigen action) in QRL.
For an arbitrary state (or action) in a quantum reinforcement
learning system, we can obtain Proposition 1.
\end{remark}

\newtheorem{proposition}{Proposition}
\begin{proposition}
An arbitrary state $|S\rangle$ (or action $|A\rangle$) in QRL can
be expanded in terms of an orthogonal set of eigen states
$|s_n\rangle$ (or eigen actions $|a_n\rangle$), i.e.
\begin{equation}
|S\rangle=\sum_{n}\alpha_n|s_{n}\rangle
\end{equation}
\begin{equation}
|A\rangle=\sum_{n}\beta_n|a_{n}\rangle
\end{equation}
where $\alpha_n$ and $\beta_n$ are probability amplitudes, and
satisfy $\sum_n|\alpha_n|^2=1$ and $\sum_n|\beta_n|^2=1$.
\end{proposition}

\begin{remark}
The states and actions in QRL are different from those in
traditional RL: (1) The sum of several states (or actions) does
not have a definite meaning in traditional RL, but the sum of
states (or actions) in QRL is still a possible state (or action)
of the same quantum system. (2) When $|S\rangle$ takes on an eigen
state $|s_n\rangle$, it is exclusive. Otherwise, it has the
probability of $|\alpha_n|^2$ to be in the eigen state
$|s_n\rangle$. The same analysis also is suitable to the action
$|A\rangle$.
\end{remark}

Since quantum computation is built upon the concept of qubit as
what has been described in Section II, for the convenience of
processing, we consider to use multiple qubit systems to express
states and actions and propose a formal representation of them for
the QRL system. Let $N_s$ and $N_a$ be the number of states and
actions, then choose numbers $m$ and $n$, which are characterized
by the following inequalities:
\begin{equation}\label{equation 20}
 N_{s}  \le 2^{m}\le 2N_{s} , N_{a}  \le 2^{n}  \le 2N_{a}
\end{equation}
And use $m$ and $n$ qubits to represent eigen state set
$S=\{|s_i\rangle\}$ and eigen action set $A=\{|a_j\rangle\}$
respectively, we can obtain the corresponding relations as
follows:
\begin{equation}
|s^{(N_{s})} \rangle  = \sum_{i = 1}^{N_{s}} C_{i} |s_{i} \rangle
\leftrightarrow |s^{(m)} \rangle  = \sum_{s = 00\cdots
0}^{\overbrace {11 \cdots 1}^{m}} C_{s} |s\rangle
\end{equation}
\begin{equation}
|a_{s_{i}}^{(N_{a})} \rangle  = \sum_{j = 1}^{N_{a}} C_{j} |a_{j}
\rangle \leftrightarrow |a_{s}^{(n)} \rangle  = \sum_{a = 00\cdots
0}^{\overbrace {11\cdots 1}^{n}} C_{a}|a\rangle
\end{equation}
In other words, the states (or actions) of a QRL system may lie in
the superposition state of eigen states (or eigen actions).
Inequalities in (\ref{equation 20}) guarantee that every states
and actions in traditional RL have corresponding representation
with eigen states and eigen actions in QRL. The probability
amplitude $C_s$ and $C_a$ are complex numbers and satisfy
\begin{equation}
\sum_{s =00\cdots 0}^{\overbrace {11 \cdots 1}^{m}} |C_{s} |^2=1
\end{equation}
\begin{equation}\label{equation 26}
\sum_{a = 00\cdots 0}^{\overbrace {11\cdots 1}^{n}} |C_{a}|^2=1
\end{equation}

\subsection{Action selection policy}
In QRL, the agent is also to learn a policy $\pi: S \times
\cup_{i\in S}A_{(i)}\to [0,1]$, which will maximize the expected
sum of discounted reward of each state. That is to say, the
mapping from states to actions is $\pi: S \to A$, and we have
\begin{equation}\label{action}
f(s)=|a_{s}^{(n)} \rangle  =  \sum_{a = 00\cdots 0}^{\overbrace
{11\cdots 1}^{n}} C_{a}|a\rangle
\end{equation}
where probability amplitude $C_a$ satisfies (\ref{equation 26}).
Here, the action selection policy is based on the collapse
postulate:

\begin{definition}[Action collapse]
When an action $|A\rangle=\sum_{n}\beta_n|a_{n}\rangle$ is
measured, it will be changed and collapse randomly into one of its
eigen actions $|a_{n}\rangle$ with the corresponding probability
$|\langle a_n|A\rangle|^{2}$:
\begin{equation}
|\langle a_n|A\rangle|^{2}=|(|a_n\rangle)^{*}|A\rangle|^{2}
=|(|a_n\rangle)^{*}\sum_{n}\beta_n|a_{n}\rangle|^{2}=|\beta_n|^2
\end{equation}
\end{definition}

\begin{remark}
According to Definition 3, when an action $|a_{s}^{(n)}\rangle$ in
(\ref{action}) is measured, we will get $|a\rangle$ with the
occurrence probability of $|C_a|^2$. In QRL algorithm, we will
amplify the probability of ``good" action according to
corresponding rewards. It is obvious that the collapse action
selection method is not a real action selection policy
theoretically. It is just a fundamental phenomenon when a quantum
state is measured, which results in a good balancing between
exploration and exploitation and a natural ``action selection"
without setting parameters. More detailed discussion about the
action selection can also be found in Ref. \cite{Chen et al 2006}
\end{remark}

\subsection{Paralleling state value updating}
In Proposition 1, we pointed out that every possible state of QRL
$|S\rangle$ can be expanded in terms of an orthogonal complete set
of eigen states $|s_n\rangle$: $|S\rangle=\sum_n
\alpha_n|s_n\rangle$. According to quantum parallelism, a certain
unitary transformation $U$ on the qubits can be implemented.
Suppose we have such an operation which can simultaneously process
these $2^m$ states with the TD(0) value updating rule
\begin{equation}
V(s)\leftarrow V(s)+\alpha(r+\gamma V(s')-V(s))
\end{equation}
where $\alpha$ is the learning rate, and the meaning of reward $r$
and value function $V$ is the same as that in traditional RL. It
is like parallel value updating of traditional RL over all states.
However, it provides an exponential-scale computation space in the
$m$-qubit linear physical space and can speed up the solutions of
related functions. In this paper we will simulate QRL process on
the traditional computer in Section V. How to realize some
specific functions of the algorithm using quantum gates in detail
is our future work.

\subsection{Probability amplitude updating}
In QRL, action selection is executed by measuring action
$|a_{s}^{(n)}\rangle$ related to certain state $|s\rangle$, which
will collapse to $|a\rangle$ with the occurrence probability of
$|C_a|^2$. So it is no doubt that probability amplitude updating
is the key of recording the ``trial-and-error" experience and
learning to be more intelligent.

As the action $|a_{s}^{(n)}\rangle$ is the superposition of $2^n$
possible eigen actions, finding out $|a\rangle$ is usually
interacting with changing its probability amplitude  for a quantum
system. The updating of probability amplitude is based on the
Grover iteration \cite{Grover 1997}. First, prepare the equally
weighted superposition of all eigen actions
\begin{equation}\label{equation 30}
|a_{0}^{(n)} \rangle  = \frac{1}{\sqrt{2^n}}(\sum_{a = 00\cdots
0}^{\overbrace {11\cdots 1}^{n}} |a\rangle)
\end{equation}
This process can be done easily by applying $n$ Hadamard gates in
sequence to $n$ independent qubits with initial states in
$|0\rangle$ respectively \cite{Grover 1997}, which can be
represented into:
\begin{equation}
H^{\otimes n}|\overbrace{00\cdots 0}^{n} \rangle  =
\frac{1}{\sqrt{2^n}}(\sum_{a = 00\cdots 0}^{\overbrace {11\cdots
1}^{n}} |a\rangle)
\end{equation}
We know that $|a\rangle$ is an eigen action, irrespective of the
value of $a$, so that
\begin{equation}
|\langle a|a_{0}^{(n)}\rangle|=\frac{1}{\sqrt{2^n}}
\end{equation}

To construct the Grover iteration we will combine two reflections
$U_a$ and $U_{a_{0}^{(n)}}$ \cite{Nielsen and Chuang 2000}
\begin{equation}
U_a=I-2|a\rangle \langle a|
\end{equation}
\begin{equation}
U_{a_{0}^{(n)}}=H^{\otimes n}(2|0\rangle\langle 0|-I)H^{\otimes
n}=2|a_{0}^{(n)}\rangle \langle a_{0}^{(n)}|-I
\end{equation}
where $I$ is unitary matrix with appropriate dimensions and
$U_{a}$ corresponds to the oracle $O$ in the Grover algorithm
\cite{Nielsen and Chuang 2000}. The external product $|a\rangle
\langle a|$ is defined $|a\rangle \langle
a|=|a\rangle(|a\rangle)^{*}$. Obviously, we have
\begin{equation}
U_a|a\rangle=(I-2|a\rangle \langle
a|)|a\rangle=|a\rangle-2|a\rangle=-|a\rangle
\end{equation}
\begin{equation}
U_a|a^{\bot}\rangle=(I-2|a\rangle \langle
a|)|a^{\bot}\rangle=|a^{\bot}\rangle
\end{equation}
where $|a^{\bot}\rangle$ represents an arbitrary state orthogonal
to $|a\rangle$. Hence $U_a$ flips the sign of the action
$|a\rangle$, but acts trivially on any action orthogonal to
$|a\rangle$. This transformation has a simple geometrical
interpretation. Acting on any vector in the $2^n$-dimensional
Hilbert space, $U_a$ reflects the vector about the hyperplane
orthogonal to $|a\rangle$. Analogous to the analysis in the Grover
algorithm, $U_a$ can be looked upon as a quantum black box, which
can effectively justify whether the action is the ``good" eigen
action. Similarly, $U_{a_{0}^{(n)}}$ preserves
$|a_{0}^{(n)}\rangle$, but flips the sign of any vector orthogonal
to $|a_{0}^{(n)}\rangle$.

Thus one Grover iteration is the unitary transformation
\cite{Chuang et al 1998}, \cite{Nielsen and Chuang 2000}
\begin{equation}
U_{Grov}=U_{a_{0}^{(n)}}U_a
\end{equation}
Now let's consider how the Grover iteration acts in the plane
spanned by $|a\rangle$ and $|a_{0}^{(n)}\rangle$. The initial
action in equation (\ref{equation 30}) can be re-expressed as
\begin{equation}
f(s)=|a_{0}^{(n)}\rangle=\frac{1}{\sqrt{2^{n}}}|a\rangle+
\sqrt{\frac{2^{n}-1}{2^{n}}} |a^{\perp}\rangle
\end{equation}

Recall that
\begin{equation}
|\langle
a_{0}^{(n)}|a\rangle|=\frac{1}{\sqrt{2^n}}\equiv\sin\theta
\end{equation}
Thus
\begin{equation}
f(s) = |a_{0}^{(n)} \rangle  = \sin\theta |a\rangle  + \cos\theta
|a^{\perp}\rangle .
\end{equation}
This procedure of Grover iteration $U_{Grov}$ can be visualized
geometrically by Fig.~1.
\begin{figure}
\centering
\includegraphics[width=3.2in]{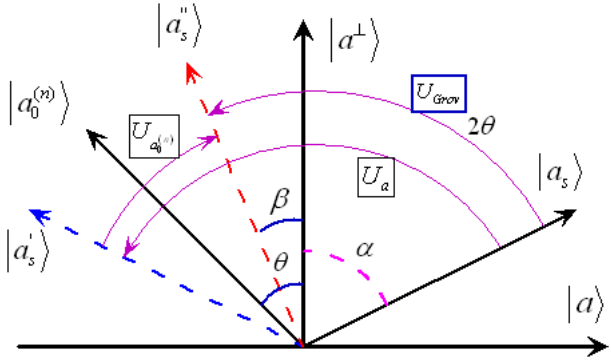}
\caption{The schematic of a single Grover iteration. $U_a$ flips
$|a_s\rangle$ into $|a_{s}'\rangle$ and $U_{a_{0}^{(n)}}$ flips
$|a_{s}'\rangle$ into $|a_{s}''\rangle$. One Grover iteration
$U_{Grov}$ rotates $|a_s\rangle$ by $2\theta$.}\label{Grover}
\end{figure}

This figure shows that $|a_{0}^{(n)}\rangle$ is rotated by
$\theta$ from the axis $|a^{\bot}\rangle$ normal to $|a\rangle$ in
the plane. $U_a$ reflects a vector $|a_s\rangle$ in the plane
about the axis $|a^{\bot}\rangle$ to $|a_{s}'\rangle$, and
$U_{a_{0}^{(n)}}$ reflects the vector $|a_{s}'\rangle$ about the
axis $|a_{0}^{(n)}\rangle$ to $|a_{s}''\rangle$. From Fig. 1 we
know that
\begin{equation}
\frac{\alpha-\beta}{2}+\beta=\theta
\end{equation}
Thus we have $\alpha+\beta=2\theta$. So one Grover iteration
$U_{Grov}=U_{a_{0}^{(n)}}U_a$ rotates any vector $|a_s\rangle$ by
$2\theta$.

We now can carry out a certain times of Grover iterations to
update the probability amplitudes according to respective rewards
and value functions. It is obvious that $2\theta$ is the updating
stepsize. Thus when an action $|a\rangle$ is executed, the
probability amplitude of $|a_{s}^{(n)}\rangle$ is updated by
carrying out $L=\text{int}(k(r+V(s')))$ times of Grover
iterations, where $\text{int}(x)$ returns the integer part of $x$.
$k$ is a parameter which indicates that the times $L$ of
iterations is proportional to $r+V(s')$. The selection of its
value is experiential in this paper and its optimization is an
open question. The probability amplitudes will be normalized with
$\sum_a|C_a|^2=1$ after each updating. According to Ref.
\cite{Boyer et al 1998}, we know that applying Grover iteration
$U_{Grov}$ for $L$ times on $|a_{0}^{(n)}\rangle $ can be
represented as
\begin{equation}
U_{Grov}^L |a_{0}^{(n)} \rangle =\sin [(2L+1) \theta]| a\rangle
+\cos [(2L+1) \theta] |a^{\perp}\rangle
\end{equation}
Obviously, we can reinforce the action $|a\rangle$ from
probability $\frac{1}{2^n}$ to $\sin^2[(2L+1)\theta]$ through
Grover iterations. Since $\sin(2L+1)\theta$ is a periodical
function about $(2L+1)\theta$ and too much iterations may also
cause small probability $\sin^2[(2L+1)\theta]$, we further select
$L=\min\{\text{int}(k(r+V(s'))),\text{int}(\frac{\pi}{4\theta}-\frac{1}{2})\}$.

\begin{remark}
The probability amplitude updating is inspired by the Grover
algorithm and the two procedures use the same amplitude
amplification technique as a subroutine. Here we want to emphasize
the difference between the probability amplitude updating and
Grover's database searching algorithm. The objective of Grover
algorithm is to search $|a\rangle$ by amplifying its occurrence
probability to almost 1, however, the aim of probability amplitude
updating process in QRL just appropriately updates (amplifies or
shrinks) corresponding amplitudes for ``good" or ``bad" eigen
actions. So the essential difference is in the times $L$ of
iterations and this can be demonstrated by Fig.~2.

\begin{figure}
\centering
\includegraphics[width=3.0in]{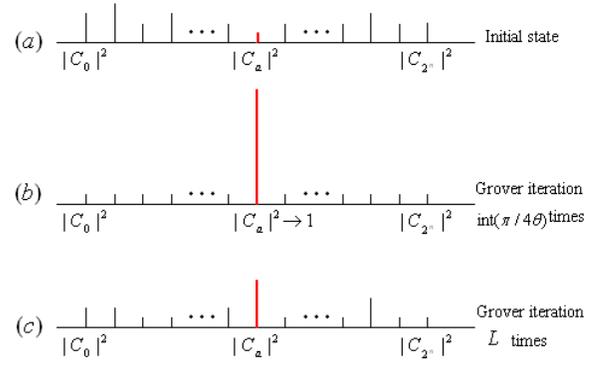}
\caption{The effect of Grover iterations in Grover algorithm and
QRL. (a) Initial state; (b) Grover iterations for amplifying
$|C_a|^2$ to almost 1; (c) Grover iterations for reinforcing
action $|a\rangle$ to probability $\sin^2[(2L+1)\theta]$}
\label{Grover-QRL}
\end{figure}
\end{remark}

\subsection{QRL algorithm}
\begin{figure*}
\centering
\includegraphics[width=4.8in]{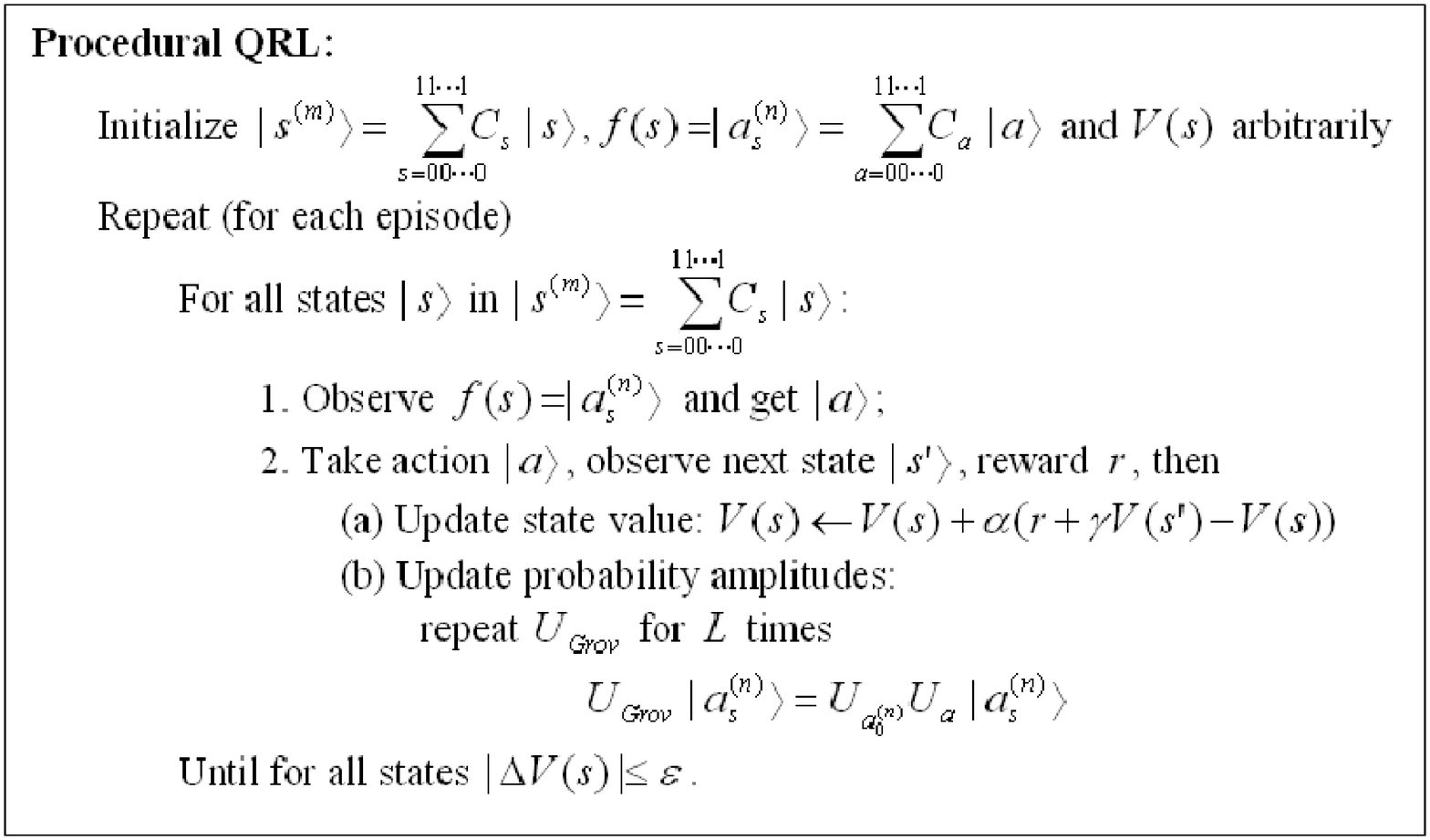}
\caption{The algorithm of a standard quantum reinforcement
learning (QRL)} \label{QRL}
\end{figure*}
Based on the above discussion, the procedural form of a standard
QRL algorithm is described as Fig.~3. In QRL algorithm, after
initializing the state and action we can observe
$|a_{s}^{(n)}\rangle$ and obtain an eigen action $|a\rangle$.
Execute this action and the system can give out next state
$|s'\rangle$, reward $r$ and state value $V(s')$. $V(s)$ is
updated by TD(0) rule, and $r$ and $V(s')$ can be used to
determine the iteration times $L$. To accomplish the task in a
practical computing device, we require some basic registers for
the storage of related information. Firstly two $m$-qubit
registers are required for all eigen states and their state values
$V(s)$, respectively. Secondly every eigen state requires two
$n$-qubit registers for their respective eigen actions stored for
two times, where one $n$-qubit register stores the action
$|a_{s}^{(n)}\rangle$ to be observed and the other $n$-qubit
register also stores the same action for preventing the memory
loss associated to the action collapse. It is worth mentioning
that this does not conflict with the no-cloning theorem
\cite{Nielsen and Chuang 2000} since the action
$|a_{s}^{(n)}\rangle$ is a certain known state at each step.
Finally several simple classical registers may be required for the
reward $r$, the times $L$, and etc.

\begin{remark}
QRL is inspired by the superposition principle of quantum state
and quantum parallelism. The action set can be represented with
the quantum state and the eigen action can be obtained by randomly
observing the simulated quantum state, which will lead to state
collapse according to the quantum measurement postulate. The
occurrence probability of every eigen action is determined by its
corresponding probability amplitude, which is updated according to
rewards and value functions. So this approach represents the whole
state-action space with the superposition of quantum state and
makes a good tradeoff between exploration and exploitation using
probability.
\end{remark}

\begin{remark}
The merit of QRL is dual. First, as for simulation algorithm on
the traditional computer it is an effective algorithm with novel
representation and computation methods. Second, the representation
and computation mode are consistent with quantum parallelism and
can speed up learning with quantum computers or quantum gates.
\end{remark}

\section{Analysis of QRL}
In this section, we discuss some theoretical properties of QRL
algorithms and provide some advice from the point of view of
engineering. Four major results are presented: (1) an asymptotic
convergence proposition for QRL algorithms, (2) the optimality and
stochastic algorithm, (3) good balancing between exploration and
exploitation, and (4) physical realization. From the following
analysis, it is obvious that QRL shows much better performance
than other methods when the searching space becomes very large.

\subsection{Convergence of QRL}
In QRL we use the temporal difference (TD) prediction for the
state value updating, and TD algorithm has been proved to converge
for absorbing Markov chain \cite{Sutton 1988} when the learning
rate is nonnegative and degressive. To generally consider the
convergence results of QRL, we have Proposition 2.

\begin{proposition}[Convergence of QRL]
For any Markov chain, QRL algorithms converge to the optimal state
value function $V^{*}(s)$ with probability 1 under proper
exploration policy when the following conditions hold (where
$\alpha_k$ is learning rate and nonnegative):
\begin{equation}\label{equation 40}
\lim_{T\to \infty}\sum_{k=1}^{T}\alpha_k=\infty, \ \ \ \
\lim_{T\to \infty}\sum_{k=1}^{T}\alpha_{k}^2<\infty
\end{equation}
\end{proposition}

\ \newline

\begin{proof}(sketch)
Based on the above analysis, QRL is a stochastic iterative
algorithm. Bertsekas and Tsitsiklis have verified the convergence
of stochastic iterative algorithms \cite{Bertsekas and Tsitsiklis
1996} when (\ref{equation 40}) holds. In fact many traditional RL
algorithms have been proved to be stochastic iterative algorithms
\cite{Bertsekas and Tsitsiklis 1996}, \cite{Sutton 1988},
\cite{Even-Dar and Mansour 2003} and QRL is the same as
traditional RL, and main differences lie in:

(1) Exploration policy is based on the collapse postulate of
quantum measurement while being observed;

(2) This kind of algorithms is carried out by quantum parallelism,
which means we update all states simultaneously and QRL is a
synchronous learning algorithm.

So the modification of RL does not affect the characteristic of
convergence and QRL algorithm converges when (\ref{equation 40})
holds.
\end{proof}

\subsection{Optimality and stochastic algorithm}
Most quantum algorithms are stochastic algorithms which can give
the correct decision-making with probability 1-$\epsilon$
($\epsilon > 0$, close to 0) after several times of repeated
computing \cite{Shor 1994}, \cite{Grover 1996}. As for quantum
reinforcement learning algorithms, optimal policies are acquired
by the collapse of quantum system and we will analyze the
optimality of these policies from two aspects as follows.

\subsubsection{QRL implemented by real quantum apparatuses}
When QRL algorithms are implemented by real quantum apparatuses,
the agent's strategy is given by the collapse of corresponding
quantum system according to probability amplitude. QRL algorithms
can not guarantee the optimality of every strategy, but it can
give the optimal decision-making with the probability
approximating to 1 by repeating computation several times. Suppose
that the agent gives an optimal strategy with the probability
$1-\epsilon$ after the agent has well learned (state value
function converges to $V^{*}(s)$). For $\epsilon \in (0,1)$, the
error probability is $\epsilon^d$ by repeating $d$ times. Hence
the agent will give the optimal strategy with the probability of
$1-\epsilon^d$ by repeating the computation for $d$ times. The QRL
algorithms on real quantum apparatuses are still effective due to
the powerful computing capability of quantum system. Our current
work has been focused on simulating QRL algorithms on the
traditional computer which also bear the characteristics inspired
by quantum systems.

\subsubsection{Simulating QRL on the traditional computer}
As mentioned above, in this paper most work has been done to
develop this kind of novel QRL algorithms by simulating on the
traditional computer. But in traditional RL theory, researchers
have argued that even if we have a complete and accurate model of
the environment's dynamics, it is usually not possible to simply
compute an optimal policy by solving the Bellman optimality
equation \cite{Sutton and Barto 1998}. What's the fact about QRL?
In QRL, the optimal value functions and optimal policies are
defined in the same way as traditional RL. The difference lies in
the representation and computing mode. The policy is probabilistic
instead of being definite using probability amplitude, which makes
it more effective and safer. But it is still obvious that
simulating QRL on the traditional computer can not speed up
learning in exponential scale since the quantum parallelism is not
really executed through real physical systems. What's more, when
more powerful computation is available, the agent will learn much
better. Then we may fall back on physical realization of quantum
computation again.

\subsection{Balancing between exploration and exploitation}
One widely used action selection scheme is
$\epsilon$\emph{-greedy} \cite{Dahl et al 2004}, \cite{Vermorel
and Mohri 2005}, where the best action is selected with
probability ($1-\epsilon$) and a random action is selected with
probability $\epsilon$ ( $\epsilon \in (0,1)$ ). The exploration
probability $\epsilon$ can be reduced over time, which moves the
agent from exploration to exploitation. The $\epsilon$-greedy
method is simple and effective but it has one drawback that when
it explores it chooses equally among all actions. This means that
it makes no difference to choose the worst action or the
next-to-best action. Another problem is that it is difficult to
choose a proper parameter $\epsilon$ which can offer an optimal
balancing between exploration and exploitation.

Another kind of action selection scheme is Boltzmann exploration
(including Softmax action selection method) \cite{Sutton and Barto
1998}, \cite{Dahl et al 2004}, \cite{Vermorel and Mohri 2005}. It
uses a positive parameter $\tau$ called the \emph{temperature} and
chooses action with the probability proportional to $e^{Q(s, \
a)/\tau}$. It can move from exploration to exploitation by
adjusting the ``temperature" parameter $\tau$. It is natural to
sample actions according to this distribution, but it is very
difficult to set and adjust a good parameter $\tau$. There are
also similar problems with simulated annealing (SA) methods
\cite{Guo et al 2004}.

We have introduced the action selecting strategy of QRL in Section
III, which is called collapse action selection method. The agent
does not bother about selecting a proper action consciously. The
action selecting process is just accomplished by the fundamental
phenomenon that it will naturally collapse to an eigen action when
an action (represented by quantum superposition state) is
measured. In the learning process, the agent can explore more
effectively since the state and action can lie in the
superposition state through parallel updating. When an action is
observed, it will collapse to an eigen action with a certain
probability. Hence QRL algorithm is essentially a kind of
probability algorithm. However, it is greatly different from
classical probability since classical algorithms forever exclude
each other for many results, but in QRL algorithm it is possible
for many results to interfere with each other to yield some global
information through some specific quantum gates such as Hadmard
gates. Compared with other exploration strategy, this mechanism
leads to a better balancing between exploration and exploitation.

In this paper, the simulated results will show that the action
selection method using the collapse phenomenon is very
extraordinary and effective. More important, it is consistent with
the physical quantum system, which makes it more natural, and the
mechanism of QRL has the potential to be implemented by real
quantum systems.

\subsection{Physical realization}
As a quantum algorithm, the physical realization of QRL is also
feasible since the two main operations occur in preparing the
equally weighted superposition state for initializing the quantum
system and carrying out a certain times of Grover iterations for
updating probability amplitude according to rewards and value
functions. These are the same operations needed in the Grover
algorithm. They can be accomplished using different combinations
of Hadamard gates and phase gates. So the physical realization of
QRL has no difficulty in principle. Moreover, the experimental
implementations of the Grover algorithm also demonstrate the
feasibility for the physical realization of our QRL algorithm.

\section{Experiments}
To evaluate QRL algorithm in practice, consider the typical
gridworld example. The gridworld environment is as shown in Fig.~4
and each cell of the grid corresponds to an individual state
(eigen state) of the environments. From any state the agent can
perform one of four primary actions (eigen actions): up, down,
left and right, and actions that would lead into a blocked cell
are not executed. The task of the algorithms is to find an optimal
policy which will let the agent move from start point $S$ to goal
point $G$ with minimized cost (or maximized rewards). An episode
is defined as one time of learning process when the agent moves
from the start state to the goal state. But when the agent cannot
find the goal state in a maximum steps (or a period of time), this
episode will be terminated and start another episode from the
start state again. So when the agent finds an optimal policy
through learning, the number of moving steps for one episode will
reduce to a minimum one.

\subsection{Experimental set-up}
In this $20\times 20 \ (0\sim19)$ gridworld, the initial state $S$
and the goal state $G$ is cell(1,1) and cell(18,18) and before
learning the agent has no information about the environment at
all. Once the agent finds the goal state it receives a reward of
$r=100$ and then ends this episode. All steps are punished by a
reward of $r=-1$. The discount factor $\gamma$ is set to 0.99 and
all of the state values V(s) are initialized as $V=0$ for all the
algorithms that we have carried out. In the first experiment, we
compare QRL algorithm with TD(0) and we also demonstrate the
expected result on a quantum computer theoretically. In the second
experiment, we give out some results of QRL algorithm with
different learning rates. For the action selection policy of TD
algorithm, we use $\epsilon$-greedy policy ($\epsilon=0.01$), that
is to say, the agent executes the ``good" action with probability
$1-\epsilon$ and chooses other actions with an equal probability.
As for QRL, the action selecting policy is obviously different
from traditional RL algorithms, which is inspired by the collapse
postulate of quantum measurement. The value of $|C_a|^2$ is used
to denote the probability of an action defined as
$f(s)=|a_{s}^{(n)}\rangle=\sum_{a=00\cdots 0}^{11\cdots
1}C_a|a\rangle$. For the four cell-to-cell actions, i.e. four
eigen actions up, down, left and right, $|C_a|^2$ is initialized
uniformly.

\begin{figure}
\centering
\includegraphics[width=3.2in]{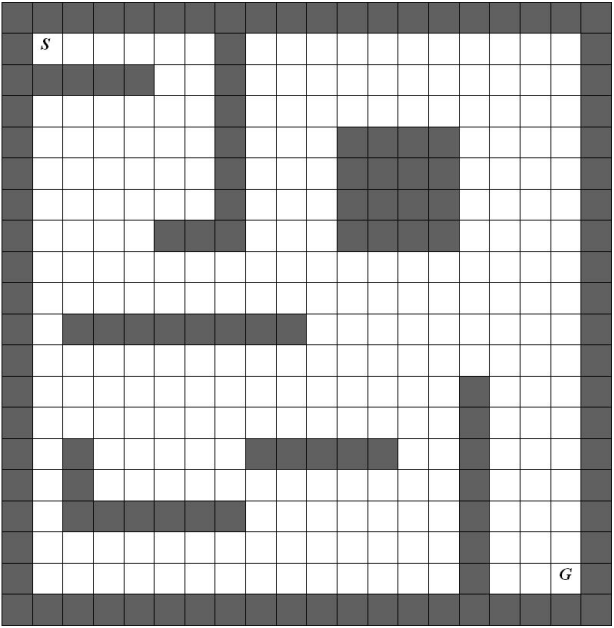}
\caption{The example is a gridworld environment with cell-to-cell
actions (up, down, left and right). The labels S and G indicate
the initial state and the goal in the simulated experiment
described in the text.} \label{grid20}
\end{figure}

\subsection{Experimental results and analysis}
Learning performance for QRL algorithm compared with TD algorithm
in traditional RL is plotted in Fig.~5, where the cases with the
good performance are chosen for both of the QRL and TD algorithms.
As shown in Fig.~5, the good cases in this gridworld example are
respectively TD algorithm with the learning rate of $\alpha=0.01$
and QRL algorithm with $\alpha=0.06$. The horizontal axis
represents the episode in the learning process and the number of
steps required is correspondingly described by the vertical
coordinate. We observe that QRL algorithm is also an effective
algorithm on the traditional computer although it is inspired by
the quantum mechanical system and is designed for quantum
computers in the future. For their respective rather good cases in
Fig.~5, QRL explores more than TD algorithm at the beginning of
learning phase, but it learns much faster and guarantees a better
balancing between exploration and exploitation. In addition, it is
much easier to tune the parameters for QRL algorithms than for
traditional ones. If the real quantum parallelism is used, we can
obtain the estimated theoretical results. What's more important,
according to the estimated theoretical results, QRL has great
potential of powerful computation provided that the quantum
computer (or related quantum apparatuses) is available in the
future, which will lead to a more effective approach for the
existing problems of learning in complex unknown environments.

\begin{figure*}
\centering
\includegraphics[width= 6.2in]{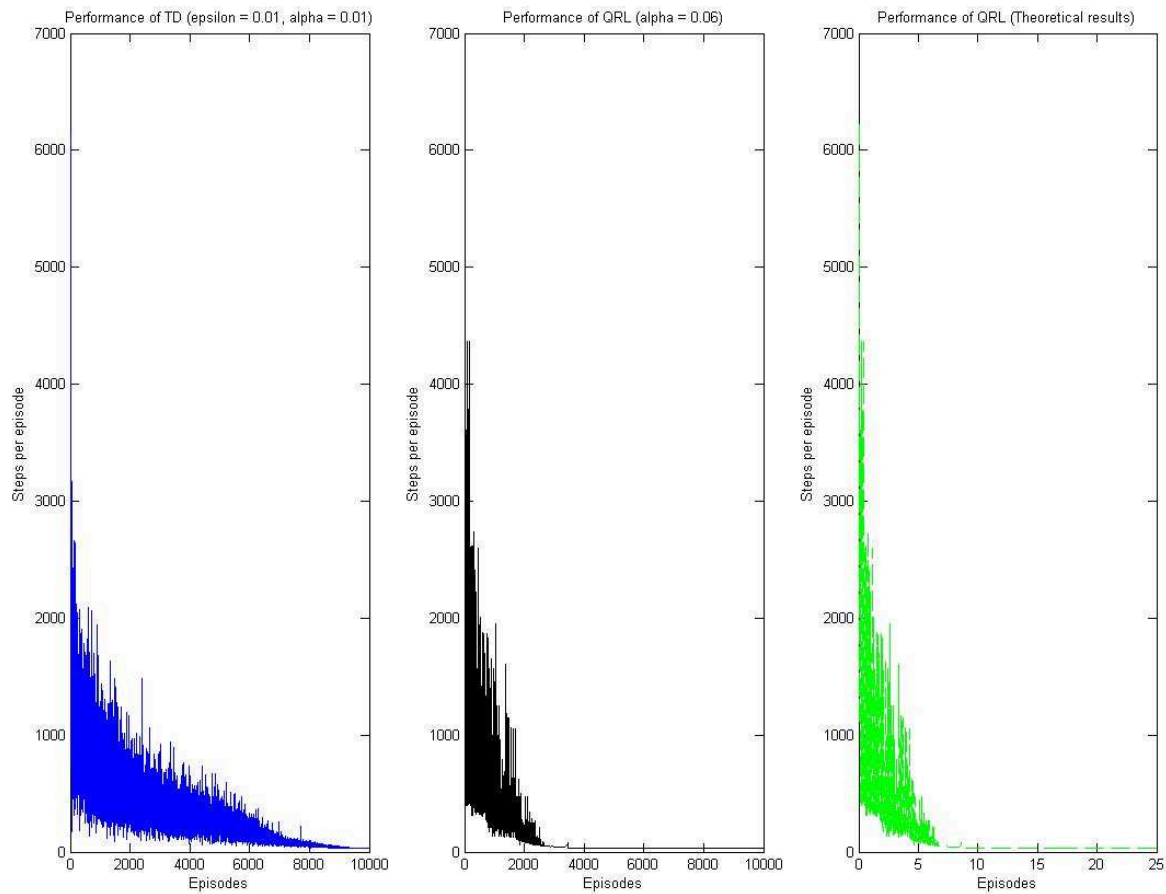}
\caption{Performance of QRL in the example of a gridworld
environment compared with TD algorithm ($\epsilon$-greedy policy)
for their respective good cases, and the expected theoretical
result on a quantum computer is also demonstrated.}
\label{QRLcomparison}
\end{figure*}

Furthermore, in the following comparison experiments we give the
results of TD(0) algorithm in QRL and RL algorithms with different
learning rates, respectively. In Fig.~6 it illustrates the results
of QRL algorithms with different learning rates: $\alpha$ (alpha),
from 0.01 to 0.11, and to give a particular description of the
learning process, we record every learning episodes. From these
figures, it can been concluded that given a proper learning rate
($0.02\le$ alpha $\le 0.10$) this algorithm learns fast and
explores much at the beginning phase, and then steadily converges
to the optimal policy that costs 36 steps to the goal state $G$.
As the learning rate increases from 0.02 to 0.09, this algorithm
learns faster. When the learning rate is 0.01 or smaller, it
explores more but learns very slow, so the learning process
converges very slowly. Compared with the result of TD in Fig.~5,
we find that the simulation result of QRL on the classical
computer does not show advantageous when the learning rate is
small (alpha $\le 0.01$). On the other hand, when the learning
rate is 0.11 or above, it cannot converge to the optimal policy
because it vibrates with too large learning rate when the policy
is near the optimal policy. Fig.~7 shows the performance of TD(0)
algorithm, and we can see that the learning process converges with
the learning rate of 0.01. But when the learning rate is bigger
(alpha=0.02, 0.03 or bigger), it becomes very hard for us to make
it converge to the optimal policy within 10000 episodes. Anyway
from Fig.~6 and Fig.~7, we can see that the convergence range of
QRL algorithm is much larger than that of traditional TD(0)
algorithm.

\begin{figure*}
\centerline {\includegraphics[width=6.2in]{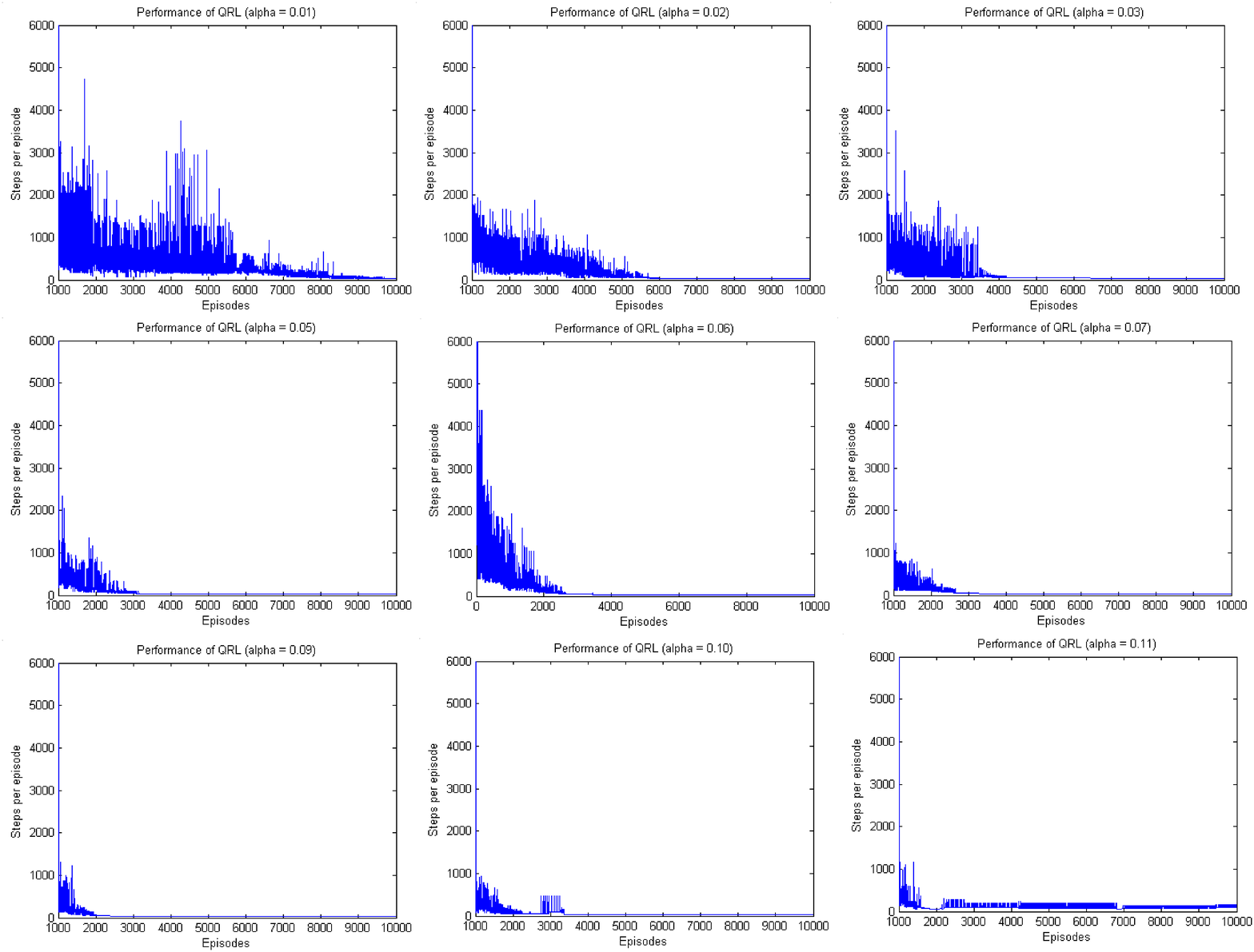}}
\caption{Comparison of QRL algorithms with different learning
rates (alpha$=0.01\sim0.11$).} \label{qrldifferentrate}
\end{figure*}

\begin{figure*}
\centerline {\includegraphics[width=5.5in]{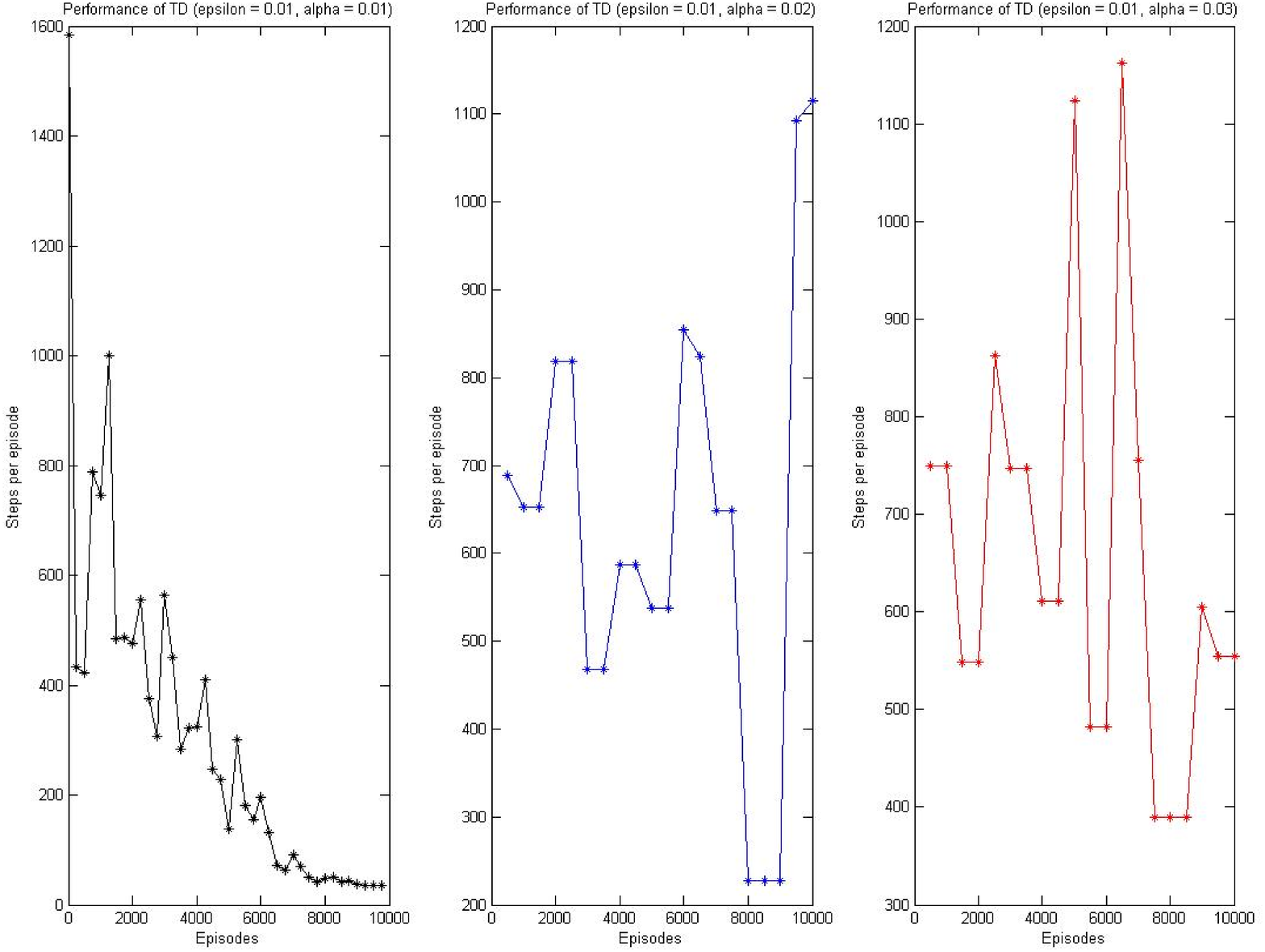}}
\caption{Comparison of TD(0) algorithms with different learning
rates (alpha$=$0.01, 0.02, 0.03).} \label{tddifferentrate}
\end{figure*}

All the results show that QRL algorithm is effective and excels
traditional RL algorithms in the following three main aspects: (1)
Action selecting policy makes a good tradeoff between exploration
and exploitation using probability, which speeds up the learning
and guarantees the searching over the whole state-action space as
well. (2) Representation is based on the superposition principle
of quantum mechanics and the updating process is carried out
through quantum parallelism, which will be much more prominent in
the future when practical quantum apparatus comes into use instead
of being simulated on the traditional computers. (3) Compared with
the experimental results in Ref. \cite{Dong et al 2006-1}, where
the simulation environment is a $13\times 13 \ (0\sim12)$
gridworld, we can see that when the state space is getting larger,
the performance of QRL is getting better than traditional RL in
simulated experiments.

\section{Discussion}
The key contribution of this paper is a novel reinforcement
learning framework called quantum reinforcement learning that
integrates quantum mechanics characteristics and reinforcement
learning theories. In this section some associated problems of QRL
on the traditional computer are discussed and some future work
regarded as important is also pointed out.

Although it is a long way for implementing such complicated
quantum systems as QRL by physical quantum systems, the simulated
version of QRL on the traditional computer has been proved
effective and also excels standard RL methods in several aspects.
To improve this approach some issues of future work is laid out as
follows, which we deem to be important.

\begin{itemize}
\item \textbf{Model of environments} An appropriate model of the
environment will make problem-solving much easier and more
efficient. This is true for most of the RL algorithms. However, to
model environments accurately and simply is a tradeoff problem. As
for QRL, this problem should be considered slightly differently
due to some of its specialities.

\item \textbf{Representations} The representations for QRL
algorithm according to different kinds of problems would be
naturally of interest ones when a learning system is designed. In
this paper, we mainly discuss problems with discrete states and
actions and a natural question is how to extend QRL to the
problems with continuous states and actions effectively.

\item \textbf{Function approximation and generalization}
Generalization is necessary for RL systems to be applied to
artificial intelligence and most engineering applications.
Function approximation is an important approach to acquire
generalization. As for QRL, this issue will be a rather
challenging task and function approximation should be considered
with the special computation mode of QRL.

\item \textbf{Theory} QRL is a new learning framework that is
different from standard RL in several aspects, such as
representation, action selection, exploration policy, updating
style, etc. So there is a lot of theoretical work to do to take
most advantage of it, especially to analyze the complexity of the
QRL algorithm and improve its representation and computation.

\item \textbf{More applications} Besides more theoretical
research, a tremendous opportunity to apply QRL algorithms to a
range of problems is needed to testify and improve this kind of
learning algorithms, especially in unknown probabilistic
environments and large learning space.
\end{itemize}

Anyway we strongly believe that QRL approaches and related
techniques will be promising for agent learning in large scale
unknown environment. This new idea of applying quantum
characteristics will also inspire the research in the area of
machine learning.

\section{Concluding Remarks}
In this paper, QRL is proposed based on the concepts and theories
of quantum computation in the light of the existing problems in RL
algorithms such as tradeoff between exploration and exploitation,
low learning speed, etc. Inspired by state superposition
principle, we introduce a framework of value updating algorithm.
The state (action) in traditional RL is looked upon as the eigen
state (eigen action) in QRL. The state (action) set can be
represented by the quantum superposition state and the eigen state
(eigen action) can be obtained by randomly observing the simulated
quantum state according to the collapse postulate of quantum
measurement. The probability of eigen state (eigen action) is
determined by the probability amplitude, which is updated
according to rewards and value functions. So it makes a good
tradeoff between exploration and exploitation and can speed up
learning as well. At the same time this novel idea will promote
related theoretical and technical research.

On the theoretical side, it gives us more inspiration to look for
new paradigms of machine learning to acquire better performance.
It also introduces the latest development of fundamental science,
such as physics and mathematics, to the area of artificial
intelligence and promotes the development of those subjects as
well. Especially the representation and essence of quantum
computation are different from classical computation and many
aspects of quantum computation are likely to evolve. Sooner or
later machine learning will also be profoundly influenced by
quantum computation theory. We have demonstrated the applicability
of quantum computation to machine learning and more interesting
results are expected in the near future.

On the technical side, the results of simulated experiments
demonstrate the feasibility of this algorithm and show its
superiority for the learning problems with huge state spaces in
unknown probabilistic environments. With the progress of quantum
technology, some fundamental quantum operations are being realized
via nuclear magnetic resonance, quantum optics, cavity-QED and ion
trap. Since the physical realization of QRL mainly needs Hadamard
gates and phase gates and both of them are relatively easy to be
implemented in quantum computation, our work also presents a new
task to implement QRL using practical quantum systems for quantum
computation and will simultaneously promote related experimental
research \cite{Dong et al 2006-1}. Once QRL becomes realizable on
real physical systems, it can be effectively used to quantum robot
learning for accomplishing some significant tasks \cite{Dong et al
2006-2}, \cite{Benioff 1998}.

Quantum computation and machine learning are both the study of the
information processing tasks. The two research fields have rapidly
grown so that it gives birth to the combining of traditional
learning algorithms and quantum computation methods, which will
influence representation and learning mechanism, and many
difficult problems could be solved appropriately in a new way.
Moreover, this idea also pioneers a new field for quantum
computation and artificial intelligence \cite{Dong et al 2006-2},
\cite{Benioff 1998}, and some efficient applications or hidden
advantages of quantum computation are probably approached from the
angle of learning and intelligence.

\section*{Acknowledgment}
The authors would like to thank two anonymous reviewers, Dr. Bo Qi
and the Associate Editor of IEEE Trans. SMCB for constructive
comments and suggestions which help clarify several concepts in our
original manuscript and have greatly improved this paper. D. Dong
also wishes to thank Prof. Lei Guo and Dr. Zairong Xi for helpful
discussions.

\end{document}